\documentclass[12pt]{elsarticle}

\usepackage{amssymb}

\usepackage{graphicx,color}
\usepackage[dvipsnames,svgnames,x11names]{xcolor}
\usepackage{textcomp}
\usepackage{amssymb,amsmath,amsfonts}
\usepackage{url}
\usepackage{xcolor}
\usepackage{a4wide}
\usepackage[version=3,arrows=pgf-filled ]{mhchem}
\usepackage[normalem]{ulem}
\usepackage{amsthm}

\usepackage{lineno}
\usepackage[capitalise]{cleveref}
\usepackage{pdflscape}

\journal{Advances in Water Resources}

\newcommand{\Dumux}{{Du\-Mu$^\text{x}$}~}
\newcommand{\Dumuxohne}{{Du\-Mu$^\text{x}$}}
\newcommand{\STOKES}{Stokes-VA~}
\newcommand{\STOKESohne}{Stokes-VA}
\newcommand{\fdrag}{$\mathbf{f}_{drag}$~}

\newtheorem*{remark}{Remark}

\begin{document}
\begin{frontmatter}

\title{Rigorous vertical averaging for pseudo-3D modeling of precipitation obstructed fluid flow}

\affiliation[inst1]{organization={Institute of Applied Mechanics (CE), University of Stuttgart},%
            addressline={Pfaffenwaldring 7},
            city={Stuttgart},
            postcode={70569},
            country={Germany}}

\affiliation[inst2]{organization={Department of Hydromechanics and Modelling of Hydrosystems, University of Stuttgart},%
            addressline={Pfaffenwaldring 61},
            city={Stuttgart},
            postcode={70569},
            country={Germany}}

\affiliation[inst3]{organization={Stuttgart Center for Simulation Science, University of Stuttgart},%
            addressline={Pfaffenwaldring 5A},
            city={Stuttgart},
            postcode={70569},
            country={Germany}}

\affiliation[inst4]{organization={Department Technical Biogeochemistry, Helmholtz Centre for Environmental Research},%
            addressline={Permoserstraße 15},
            city={Leipzig},
            postcode={04318},
            country={Germany}}

\affiliation[inst5]{organization={Department of Chemical and Biological Engineering, Center for Biofilm Engineering, Montana State University},%
            addressline={366 Barnard Hall},
            city={Bozeman},
            postcode={MT 59717},
            country={USA}}

\author[inst1,inst3]{David Krach\textsuperscript{$\ast\,$} } %
\author[inst4,inst5]{Felix Weinhardt\textsuperscript{$\ast\,$} } %
\author[inst2]{Holger Class} %
\author[inst1,inst3]{Holger Steeb} %
\author[inst2,inst3]{Martin Schneider} %

\begin{abstract}
Computing effective hydraulic properties of porous materials with a dynamically changing pore morphology presents a complex and critical challenge in both theoretical and applied sciences. This paper presents a tool to effectively compute the intrinsic permeability $\mathbf{k}(\mathbf{x}, t)$ for domains with both temporally and spatially varying pore structures, by utilizing a sample of snapshots of microfluidic experiments with calcium carbonate precipitation. To compute large domains at reasonable cost, a pseudo-3D Stokes solver with an additional viscous drag term is employed. In order to simulate the intricate domains with highly heterogeneous pore space, we derive vertically averaged governing equations including a vertically averaged drag term from scratch and without limiting assumptions. For exemplary precipitates (semi-spheres) of varying sizes within a rectangular channel with fixed solid boundaries, results are compared across multiple modeling approaches, each employing different drag force formulations. In addition, the performance of the derived models is examined for segments of an EICP experiment in terms of accuracy (comparison with 3D solvers) and computational efficiency. Finally, an entire experimental domain is used to demonstrate the capabilities offered by the solver thanks to its efficiency as well as accuracy and the permeability tensor $\mathbf{k} = k_{ij}\, \mathbf{e}_i \otimes \mathbf{e}_j$ is computed and compared with experimental results. In all these applications, the vertically averaged drag term proves to be the optimal solution in terms of speed and accuracy, thus allowing for an efficient analysis of the non-uniformity of local fluid flow fluctuations resulting from alterations in the pore space.

\end{abstract}


\end{frontmatter}

\section{Introduction}
\label{sec:intro}
\noindent The computation of intrinsic permeability $\mathbf{k}$ of porous materials, which possess a pore structure that changes dynamically, poses a complex and significant challenge in many branches of engineering and applied sciences. The evolution of pore space, whether due to mechanical deformation, chemical dissolution and precipitation, biological activity, or thermal effects,  directly influences fluid transport properties, altering porosity, flow pathways, and effective permeability. This paper focuses on enzymatically induced calcium carbonate precipitation (EICP) and a yielding clogging process whose understanding is essential for applications ranging from enhanced oil recovery, groundwater remediation to the  production of sustainable building materials \cite{Phillipsetal2013b, Nethingetal2020}. Determining porosity-permeability relationships for representative elementary volume (REV) scale simulation models remains a work in progress \cite{Hommeletal2018, Kohlhaas2025, Tang2020}. Microfluidic experiments, e.g., \cite{weinhardt2021experimental, weinhardt2022spatiotemporal}, in combination with pore-scale simulations, have become indispensable to capture these dynamic changes and quantify their impact on hydraulic properties or even macroscopic transport behavior. We propose an approach to effectively simulate domains with both temporally and spatially varying pore structure by utilizing a sample of experimental snapshots and conducting a series of simulations (one per snapshot) to ultimately enable accurate predictions of permeability under evolving conditions.

One classic problem in pseudo-3D/2D fluid mechanics is the analysis of creeping flow between parallel plates. Traditionally, this problem has been approached under the assumption of a constant aperture between the plates \cite{Thompson1968, flekkoy1995hydrodynamic}. However, many practical applications, e.g. in fractured material,  involve geometries in which the aperture varies spatially. Such variations can significantly influence the flow characteristics, leading to complex behaviors that are not adequately captured by traditional models, which generally assume that velocities and fluid accelerations in the out-of-plane direction are not present. This assumption is violated for geometries which this study is focusing on. As a result, the production of viscous momentum dispersion, introduced as a drag term, is significantly underestimated and permeability is overestimated. There exist several approaches to address this issue, along with models tailored to diverse applications, such as single-phase flow with calcium carbonate precipitation \cite{Krach2025a}, multiphase flow \cite{Holme1992, Fernandez2001, Horgue2013}, and flow through fractures \cite{Dzikowski2018}.
Closely related is lubrication theory and its higher-order extensions \cite{Tavakol2017,Housiadas2022}, which likewise expand the Stokes equations in the aspect ratio. Its aim, however, is the resolved velocity and pressure field rather than the averaged equation, and the resulting complexity has to the best of our knowledge confined it to two-dimensional and axisymmetric geometries. Three-dimensional two-wall results exist only for small-amplitude periodic bumps \cite{Wang2004}. For arbitrary two-sided aperture variations in two horizontal dimensions, as considered here, no vertically averaged momentum equation with a rigorously derived drag term is, to the best of our knowledge, available.

\noindent In a previous study \cite{Krach2025a}, we have systematically investigated the applicability of classical pseudo-3D Stokes equations to model fluid flow in porous media with varying apertures. Therein, the governing equations are tailored by a data-driven extension $\boldsymbol{\lambda}$ to the drag term $\mathbf{f}_{drag}$ to reflect the intricate characteristics of the local pore geometry and the resulting momentum dispersion. This data-driven approach has proven highly effective, yielding results that closely align with fully resolved high resolution 3D simulations. However, one disadvantage of the data-driven approach is that preprocessing requires approximately the same amount of computation time as solving the equations subsequently. In addition, this formulation has an inherent error that arises from the fact that the plain 2D Stokes equations are solved instead of the actual vertically averaged Stokes equations. Using 2D Stokes equations requires an indiscriminate averaging to be involved in the preprocessing for $\boldsymbol{\lambda}$ which e.g., shifts local velocity maxima. Recognizing this, we seize the opportunity and propose an advanced methodological framework based on a mathematically rigorous vertical averaging process for the governing equations and the drag term $\mathbf{f}_{drag}$ designed to be universally applicable and reduce the reliance on empirical adjustments.
To further elevate the accuracy of our model, we upgraded the vertically averaged drag term with a refined data-driven approach.

This study addresses the rigorous derivation, implementation, and analysis of vertically averaged equations for fluid flow in porous media, with a focus on overcoming limitations imposed by traditional assumptions/requirements to the pore space. The objective is threefold.
First, the experimental methodologies (Section~\ref{ssec:experimental_setup}) and data produced (Section~\ref{ssec:local_porespace}) as well as the governing equations (Section~\ref{ssec:governing_eq}) are presented. This includes a rigorous derivation of vertically averaged equations, incorporating a novel vertically averaged drag term that avoids a-priori geometric assumptions known to introduce inaccuracies (Section~\ref{ssec:fdrag_derivation}).
Next, this drag term is implemented in the pseudo-3D solver of \Dumux (DUNE for Multi-\{Phase, Component, Scale, Physics, \ldots\} flow and transport in porous media) \cite{Kochetal2020}. In addition, enhancing its precision through a hybrid approach that combines the derived vertically averaged drag term with a data-driven methodology is possible. This hybrid framework is designed to optimize results in terms of accuracy for domains with complex cross-sections.
Finally, we evaluate the performance of the proposed model across multiple scales (pore-scale to macro-scale) using benchmark cases (Section~\ref{ssec:results_singleprec}), segments of geometries (Section~\ref{ssec:results_segments}), and entire microfluidic domains (Section~\ref{ssec:results_wholedomain}). The analysis further considers computational efficiency, ensuring the model's suitability for intricate, real-world/large domain simulation campaigns.

\section{Materials and methods}
\subsection{Microfluidic setup and experiments}\label{ssec:experimental_setup}

\begin{figure}[h!]
    \centering
    \includegraphics[width=0.99\textwidth]{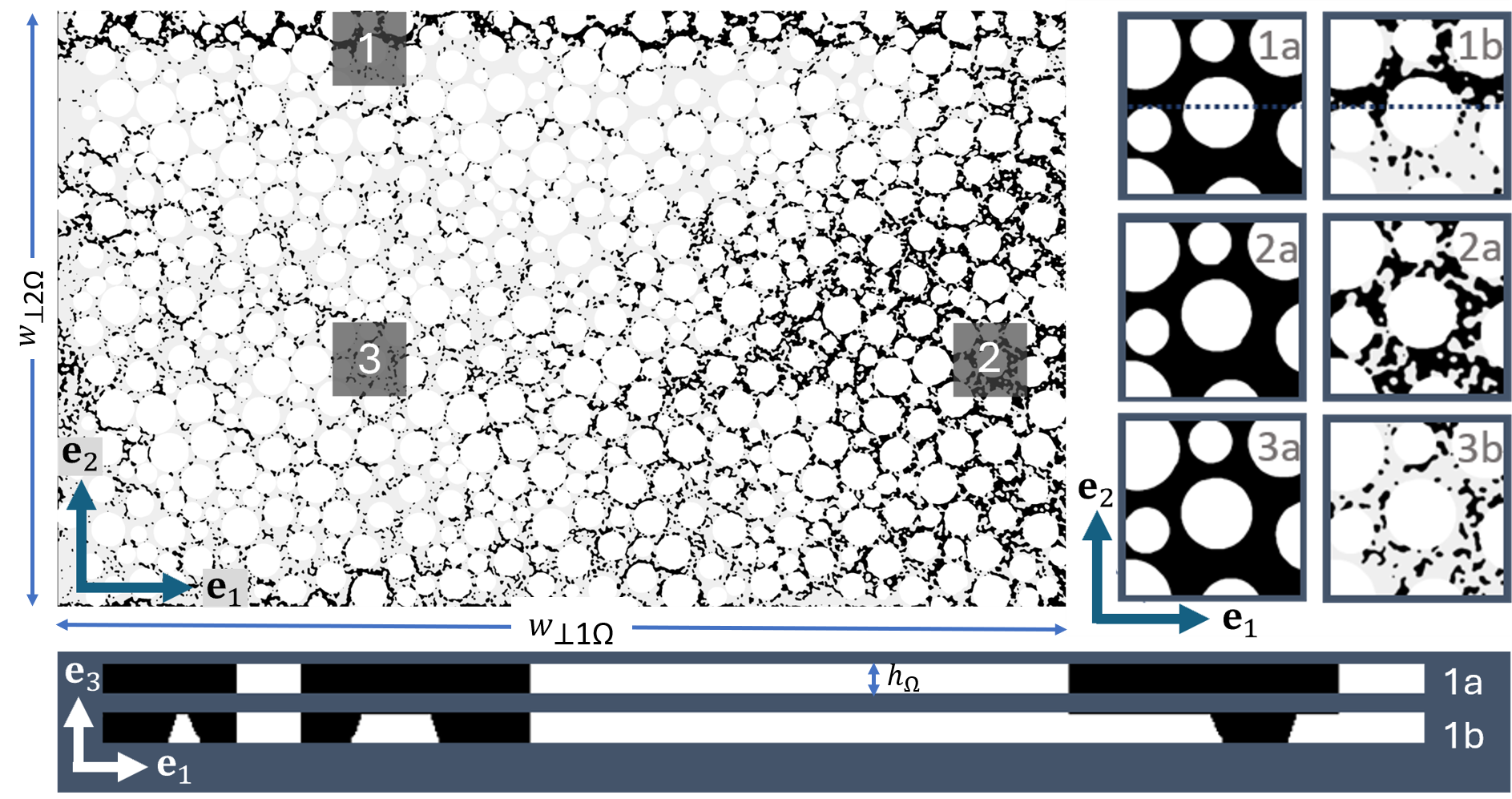}
    \caption{Left: Pore structure at the final state of the considered time interval. Black is fluid, white is solid, gray values represent the calcium carbonate precipitates. The numbered gray boxes indicate the location of the three investigated segments. Right: Enlarged representation of the three segments at initial and final state of the experiment. Bottom: Cross section of \textit{Segment 1} at initial state (1a) and final state (1b). Reprinted from \cite{Krach2025a}.}
    \label{fig::cellsegments_experiment}
\end{figure}

The \textit{Microfluidic EICP experiment} analyzed in this study builds on the dataset \textit{Quasi-2D-1} from \citet{Weinhardtetal2022, darus-1799_2022}.
The borosilicate glass microfluidic cell featured a porous domain of \(20.5 \times 11.9 \times 0.035\,\mathrm{mm}\) (\(w_{\perp2\Omega} \times w_{\perp1\Omega} \times h_\Omega\)), an initial porosity of \(\phi_0 = 0.44\), and an initial permeability of \(k_{11,0} = 3.6 \times 10^{-11}\,\mathrm{m^2}\).
Urease solution (extracted from \(5\,\mathrm{g/L} \) Jack Bean Meal and an equimolar \(\mathrm{CaCl_2}\)/urea solution of 1/3\,\(\mathrm{mol/L}\)) were co-injected at \(0.04\,\mathrm{\mu L/s}\);
the flow rate was temporarily reduced before resaturation was required after approximately 26\,h. Only the post-restart phase with a constant flow rate of \(0.04\,\mathrm{\mu L/s}\) over 4\,h was analyzed, and the inlet--outlet pressure difference was used to assess permeability evolution.
Optical microscopy revealed a persistent preferential flow path and pronounced pore-scale heterogeneity (see Figure~\ref{fig::cellsegments_experiment}).
Three representative segments were selected for numerical permeability estimation: along the main flow path (1), near the outlet with sparse homogeneous precipitation (2), and within a densely precipitated central region (3).

Image processing followed the established workflow of \citet{Weinhardtetal2022, darus-1799_2022, hommel2022effects}.
Registered and binarized 8-bit grayscale images provided 2D crystal projections, which were extruded into 3D assuming frustum-shaped precipitates and the known cell height.
Given that only 2D grayscale images are available, the most plausible hypothesis is that precipitate formation initiates at the base of the channel and progresses upward. However, this assumption cannot be definitively validated or refuted based solely on the available 2D datasets.
The reconstructed pore geometry was exported in voxel format.
It was further assumed that precipitates never fully close the cell height, leaving at least one-voxel aperture in the reconstructed cross-section (see Fig.~\ref{fig::cellsegments_experiment}).

Microfluidic models for investigating EICP (Enzymatically Induced Calcite Precipitation) and clogging \cite{weinhardt2021experimental, weinhardt2022spatiotemporal, wang2019, wang2022} naturally exhibit a three-dimensional (3D) structure, where the height $h_\Omega$ of the actual porous material is significantly smaller than the other two dimensions $w_{\perp 2 \Omega}, w_{\perp 1 \Omega} \gg h_\Omega$.

To effectively conduct simulations in a 2D domain, an averaging process is applied to the physical quantities over the height of the model $h_\Omega$ or, alternatively, over the height of the available permeable pore space $h(\mathbf{x})$ at any given spatial position $\mathbf{x} = (x_1,x_2)^T$. The local height $h(\mathbf{x})$ depends on the growth of the precipitate (it is therefore also dependent on time) at each spatial point within the pore space of the porous material.
The choice of the averaging approach is a crucial consideration, as analyzed in the following sections. In our case, the averaging is always performed along the direction of the basis vector $\mathbf{e}_3$, corresponding to the coordinate axis $x_3$.

\subsection{Local porespace quantities}\label{ssec:local_porespace}
\begin{figure}
    \centering
    \includegraphics[width=0.8\linewidth]{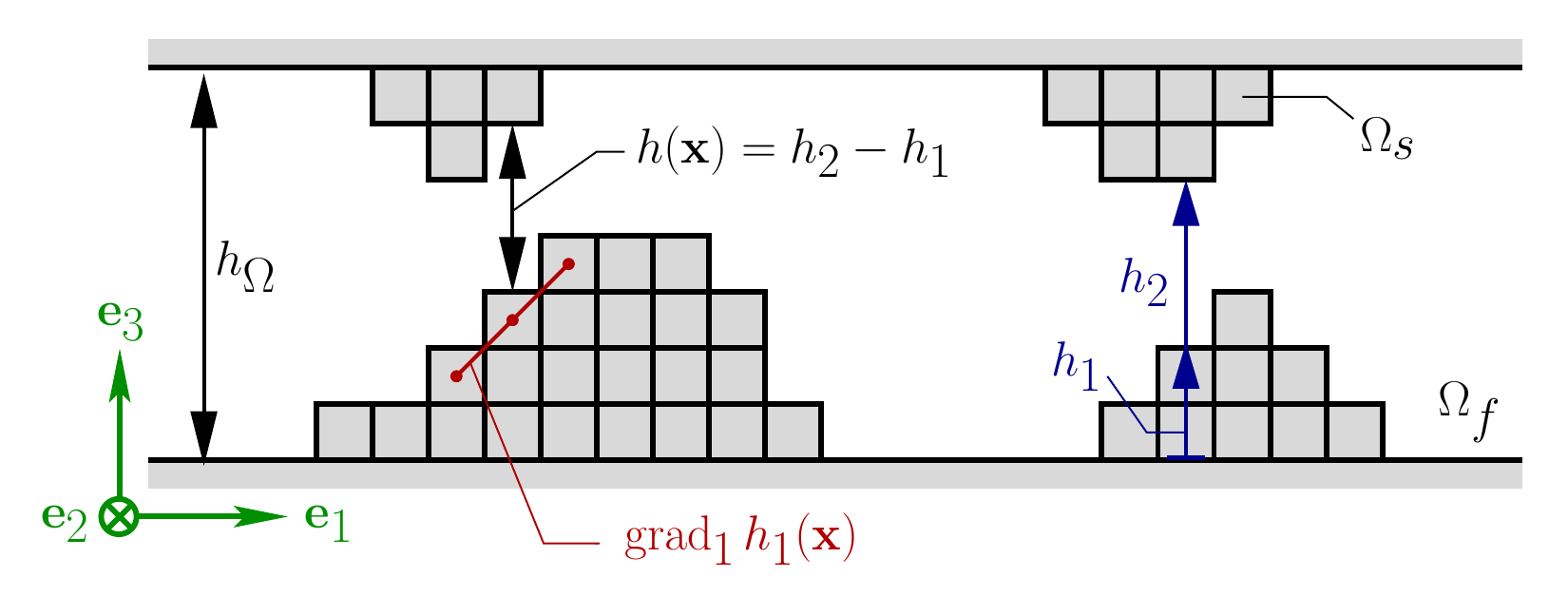}
    \caption{Schematic representation of the local height distribution $h(\mathbf{x}) = h_2 - h_1$ in a channel with an arbitrary precipitation profile. The solid or fluid phase ($\Omega_s$, $\Omega_f$) occupy the entire domain within the top and bottom plate. The local porosity gradient $\mathrm{grad}_1\, h_1(\mathbf{x})$ indicates the local slope of the solid precipitate. The $\mathbf{e}_3$-direction is collapsed and is not resolved in pseudo-3D \Dumux simulations.}
    \label{fig:porespace}
\end{figure}

\noindent The original drag term $\mathbf{f}_{drag} = - 12 \frac{\mu}{h_\Omega^2} \mathbf{v}^{2D}$ (used in \cref{eq:2d_brinkman_gov_equations}), by \citet{flekkoy1995hydrodynamic},
where $\mu$ is the dynamic viscosity and $h_\Omega$ is the domain height (both are assumed constant), depends merely on the local vertically-averaged velocity $\mathbf{v}^{2D}$. It does not account for any spatially varying features of the porespace. This assumption is valid for simple flow between parallel plates ($h(\mathbf{x}) = h_\Omega \, \forall \, \mathbf{x}$) but may not hold for flows with varying apertures, since it neglects vertical acceleration of the fluid. In cases that violate this premise, it is essential to evaluate the local pore space properties and adjust the drag term accordingly. The relevant quantities that impact the modified drag terms are illustrated in Figure~\ref{fig:porespace}.

The key parameters are the local height $h(\mathbf{x}) \in [0, h_\Omega]$, representing the space occupied by the fluid phase $\Omega_f$ between either the solid phase $\Omega_s$ (precipitate) or the bounding plates (top and/or bottom), and the local gradient of the solid phase (denoted here as $\mathrm{grad} \, h_1(\mathbf{x})$). In the absence of 3D data, both parameters are computed from grayscale values (see Figure~\ref{fig::cellsegments_experiment}).

\subsection{Governing equations}\label{ssec:governing_eq}
The basic assumptions for all, pseudo-3D as well as 3D, simulations are that (1) the pore space of the porous material is completely filled by an incompressible fluid phase, (2) the dynamic viscosity $\mu$ of this Newtonian fluid is constant, (3) the Reynolds numbers are small enough to remain within the creeping flow regime $\mathrm{Re} \leq 1$ , and (4) steady-state conditions are given. There is no gravitational force $\mathbf{b}$, which results in the following for the external source term $\rho\mathbf{b} = \boldsymbol{0}$. Given that, the balance of linear momentum is known as Stokes equations
\begin{equation}\label{eq:3d_gov_equations}
    \mathbf{0}^{3D}= \mu \, \Delta^{3D} \mathbf{v}^{3D} - \nabla^{3D}\, p^{3D} \quad \text{and} \quad \nabla^{3D} \cdot \mathbf{v}^{3D} = 0 \, ,
\end{equation}
where $\rho$, $\mathbf{v}$, and $p$  are the density, velocity, and pressure of the pore fluid. For consistency and clarity, the dimensions of all operators, field quantities, and related terms are explicitly stated throughout this paper $\left( (\bullet)^{2D}, (\bullet)^{3D}  \right)$.
We refer to equations and models whose field functions are defined in 2D Euclidian coordinate space $\mathbb{R}^2$ as pseudo-3D, if additional terms are taken into consideration, which, based on an averaging procedure, incorporate 3D effects. The most straightforward approach (used in \citet{Krach2025a}) is to merely extend the completely 2D version of the equations

\begin{equation}\label{eq:2d_gov_equations}
    \mathbf{0}^{2D}= \mu \, \Delta^{2D} \, \mathbf{v}^{2D} - \nabla^{2D}\, p^{2D} \quad \text{and} \quad \nabla^{2D} \cdot \mathbf{v}^{2D} = 0 \, ,
\end{equation}

\noindent by an additional drag term $\mathbf{f}_{drag}$ accounting for the viscous drag or in other words the resistance that the fluid experiences due to the presence of top and bottom solid surfaces, yielding:

\begin{equation}\label{eq:2d_gov_equations_with_drag}
    -\mathbf{f}_{drag} = \mu \, \Delta^{2D} \, \mathbf{v}^{2D} - \nabla^{2D}\, p^{2D} \quad \text{and} \quad \nabla^{2D} \cdot \mathbf{v}^{2D} = 0 \, .
\end{equation}

\noindent In \cref{eq:2d_gov_equations_with_drag}, $\mathbf{v}^{2D}$ and $p^{2D}$ are treated as plain 2D fields without reference to an averaging procedure. This is consistent as long as the aperture is uniform, $h(\mathbf{x}) \equiv h_\Omega$. For a spatially varying aperture, the vertically-averaged equations depend on how the 2D fields are defined. In \cref{ssec:fdrag_derivation} we derive them from the 3D Stokes equations, using a superficial average (normalized by $h_\Omega$) for the velocity and an intrinsic average (normalized by the local height $h(\mathbf{x})$) for the pressure. The resulting governing equations, which are used for the derived model throughout this work, read

\begin{equation}\label{eq:2d_brinkman_gov_equations}
    -\mathbf{f}_{drag} = \mu \, \Delta^{2D} \, \mathbf{v}^{2D} - \frac{h(\mathbf{x})}{h_\Omega} \nabla^{2D}\, p^{2D} \quad \text{and} \quad \nabla^{2D} \cdot \mathbf{v}^{2D} = 0 \, .
\end{equation}

\noindent \citet{flekkoy1995hydrodynamic} proposed the following drag term
\begin{equation}\label{eq:org_drag_term}
\mathbf{f}_{drag} = - 12 \frac{\mu}{h_\Omega^2} \mathbf{v}^{2D}
\end{equation}
which has already been used for single-phase fluid flow \cite{Boek2010,Venturolietal2006,laleian2015incompressible}, density-driven flow \citep{class2020experimental}, viscous fingering \cite{GROSFILS2003,Al-Ghoul2004}, for PEM fuel cells \cite{Nandjou2016,Tardy2019} or transport under multiphase conditions \cite{Ferrari2015,Saeibehrouzi2024a,Saeibehrouzi2024b} in micromodels. However, this approach cannot account for the effects arising from nonzero out-of-plane velocities $v_3 \neq 0$. To address this problem, an additional factor $\boldsymbol{\lambda}$ based on the local pore geometry can be introduced
\begin{equation}\label{eq:krach_drag_term}
\mathbf{f}_{drag} = - 12 \boldsymbol{\lambda} \frac{\mu}{h(\mathbf{x})^2} \mathbf{v}^{2D} \, ,
\end{equation}
as suggested in \citet{Krach2025a}. However, this empirical and data-driven approach has the disadvantage that it is tailored to a specific problem, and the calculation of $\boldsymbol{\lambda}$ is rather computationally expensive.

\subsection{Rigorous derivation of governing equations}\label{ssec:fdrag_derivation}
Within this section, vertically-averaged equations are derived with respect to the averaged variables
\begin{equation}
    v_i^{2D} := \frac{1}{h_\Omega} \int_{h_1}^{h_2}  v_i^{3D}  \, \mathrm{d}x_3, \quad p^{2D} := \frac{1}{h_2 - h_1} \int_{h_1}^{h_2}  p^{3D}  \, \mathrm{d}x_3 \, .
\end{equation}
Here, we use the so-called superficial average for the velocity and the so-called intrinsic average for pressure.

Using the Laplace operator
$\Delta^{3D}  := \nabla^{3D} \cdot \nabla^{3D} =\sum_{j=1}^3 \frac{\partial^2}{\partial x_j^2}$
and the splitting $\Delta^{3D} = \Delta^{2D} + \frac{\partial^2}{\partial x_3^2}$, \cref{eq:3d_gov_equations} can be equivalently written as
\begin{equation}\label{eq:3d_gov_equations_splitting}
    -\mu \frac{\partial^2}{\partial x_3^2} \mathbf{v}^{3D} = \mu \, \Delta^{2D} \, \mathbf{v}^{3D} - \nabla^{3D}\, p^{3D} \quad \text{and} \quad \nabla^{3D} \cdot \mathbf{v}^{3D} = 0.
\end{equation}

In the following, we will also make use of the boundary parameterizations $\mathbf{\Phi}$ \cref{eq:boundaryparametrization} to improve readability. The only assumption we make on the velocity field is a no-slip boundary condition, i.e $ \mathbf{v}^{3D}(\mathbf{\Phi}_j(\mathbf{x}^{2D})) = \mathbf{v}^{3D}(\mathbf{x}^{2D}, h_j(\mathbf{x}^{2D})) = 0$, $j = 1,2$.

\subsubsection*{Continuity equation}
Integrating the continuity equation over the vertical pore space \cref{fig:porespace}, yields
\begin{equation}
    \int_{h_1}^{h_2}  \nabla^{3D} \cdot \mathbf{v}^{3D} \, \mathrm{d}x_3 = \sum_{j=1}^3 \int_{h_1}^{h_2} \frac{\partial}{\partial x_j}  v_j^{3D} \, \mathrm{d}x_3 = \sum_{j=1}^2 \int_{h_1}^{h_2}   \frac{\partial}{\partial x_j}  v_j^{3D} \, \mathrm{d}x_3 + v^{3D}_3 \bigg|_{h_1}^{h_2}.
\end{equation}

Applying \cref{eq:div-Leibniz} and using the fact that $v^{3D}_3 \bigg|_{h_1}^{h_2} = 0$ gives
\begin{equation}
  0 = \int_{h_1}^{h_2}  \nabla^{3D} \cdot \mathbf{v}^{3D} \, \mathrm{d}x_3 =  \sum_{j=1}^2 \frac{\mathrm{d}}{ \mathrm{d} x_j}  \int_{h_1}^{h_2}  v_j^{3D} \, \mathrm{d}x_3  = \nabla^{2D} \cdot ( h_{\Omega} \mathbf{v}^{2D} ).
\end{equation}
Division by $h_{\Omega}$ gives the classical continuity equation for $\mathbf{v}^{2D}$.

\subsubsection*{Momentum equation}
Integrating the ith momentum balance \cref{eq:3d_gov_equations_splitting}, $i \in \lbrace 1,2 \rbrace$, yields
\begin{equation}
   -\mu \frac{\partial}{\partial x_3} v_i^{3D} \bigg|_{h_1}^{h_2} =  \mu  \int_{h_1}^{h_2} \Delta^{2D} \, v_i^{3D} \, \mathrm{d}x_3  -  \int_{h_1}^{h_2} \frac{\partial}{\partial x_i} p^{3D} \, \mathrm{d}x_3.
\end{equation}
Applying the Leibniz rule twice for the Laplace operator, see \cref{{eq:leibnizlaplace}}, results in
\begin{equation}
\begin{aligned}
    -\mu \frac{\partial}{\partial x_3} v_i^{3D} \bigg|_{h_1}^{h_2} + \sum_{j=1}^2 \mu \left( \frac{\partial h_2}{\partial x_j} \frac{\partial v_{i}^{3D}}{\partial x_j}(\mathbf{\Phi}_2) - \frac{\partial h_1}{\partial x_j} \frac{\partial v_{i}^{3D}}{\partial x_j}(\mathbf{\Phi}_1) \right) \\
    =  \mu  h_\Omega \Delta^{2D} v_i^{2D}  -  \int_{h_1}^{h_2} \frac{\partial}{\partial x_i} p^{3D} \, \mathrm{d}x_3.
\end{aligned}
\end{equation}
Using  \cref{eq:velderivb} to reformulate the velocity derivatives gives
\begin{equation}
    \mu \sum_{j=1}^2 (-1)^{j+1}\frac{\partial}{\partial x_3} v_i^{3D}(\mathbf{\Phi}_j) \left( 1 + \lVert \nabla h_j \rVert^2 \right)  =  \mu  h_\Omega \Delta^{2D} v_i^{2D}  -  \int_{h_1}^{h_2} \frac{\partial}{\partial x_i} p^{3D} \, \mathrm{d}x_3.
\end{equation}
Applying again the Leibniz rule for the pressure term, see \cref{eq:leibnizpressure}, yields
\begin{equation}
\begin{aligned}
    &\mu \sum_{j=1}^2 (-1)^{j+1}\frac{\partial}{\partial x_3} v_i^{3D}(\mathbf{\Phi}_j) \left( 1 + \lVert \nabla h_j \rVert^2 \right)  \\
    + & \sum_{j=1}^2 (-1)^{j+1} \frac{\partial h_j}{\partial x_i} \left( p^{3D}(\mathbf{\Phi}_j) - p^{2D} \right) =  \mu  h_\Omega \Delta^{2D} v_i^{2D}  - (h_2 - h_1) \frac{\partial}{\partial x_i}  p^{2D}.
\end{aligned}
\end{equation}
Inserting \cref{eq:pressurebc} to eliminate $p^{3D}(\mathbf{\Phi}_j)$ allows to rewrite
\begin{equation}
\begin{aligned}
    &\mu \sum_{j=1}^2 (-1)^{j+1}\frac{\partial}{\partial x_3} v_i^{3D}(\mathbf{\Phi}_j) \left( 1 + \lVert \nabla h_j \rVert^2 \right)  \\
    + &\mu \sum_{j=1}^2 (-1)^{j+1} \frac{\partial h_j}{\partial x_i} \left( \frac{\partial }{\partial x_3}\mathbf{v}^{3D}_{1:2}(\mathbf{\Phi}_j) \cdot \nabla^{2D}h_j + \lambda_j^p\right) =  \mu  h_\Omega \Delta^{2D} v_i^{2D}  - (h_2 - h_1) \frac{\partial}{\partial x_i}  p^{2D}.
\end{aligned}
\end{equation}
Grouping terms together and dividing by $h_{\Omega}$, the above equation is written in vector notation as
\begin{equation}
\begin{aligned}
    &\frac{\mu}{h_\Omega} \sum_{j=1}^2 (-1)^{j+1}  \left( (1 + \lVert \nabla^{2D} h_j \rVert^2) \mathbf{I} + \nabla^{2D} h_j \nabla^{T,2D} h_j \right) \left(\frac{\partial }{\partial x_3}\mathbf{v}^{3D}_{1:2}\right)(\mathbf{x}^{2D},h_j(\mathbf{x}^{2D}))) \\
   +&\frac{\mu}{h_\Omega} \sum_{j=1}^2 (-1)^{j+1} \lambda^p_j(\mathbf{x}^{2D}) \nabla^{2D} h_j =  \mu \Delta^{2D} \mathbf{v}^{2D}  - \frac{h_2 - h_1}{ h_\Omega} \nabla^{2D}  p^{2D}.
\end{aligned}
\end{equation}

With this, the following drag term $\mathbf{f}_{drag} = \mathbf{f}^p_{drag} + \mathbf{f}^v_{drag}$ is derived
\begin{equation}
\begin{aligned}
    -\mathbf{f}^p_{drag} &= \frac{\mu}{h_\Omega} \sum_{j=1}^2 (-1)^{j+1}  \lambda^p_j(\mathbf{x}^{2D}) \nabla^{2D} h_j, \\
    -\mathbf{f}^v_{drag} &= \frac{\mu}{h_\Omega} \sum_{j=1}^2 (-1)^{j+1} \mathbf{F}( \nabla^{2D} h_j) \left(\frac{\partial }{\partial x_3}\mathbf{v}^{3D}_{1:2}\right)(\mathbf{x}^{2D},h_j(\mathbf{x}^{2D})),
\end{aligned}
\end{equation}
with tensor
\begin{equation}
\mathbf{F}( \nabla^{2D} h_j) :=(1 + \lVert \nabla^{2D} h_j \rVert^2) \mathbf{I} + \nabla^{2D} h_j \nabla^{T,2D} h_j.
\end{equation}

Please note that so far the only assumption we made was the no-slip boundary conditions. For further reformulation, we make some additional assumptions.
For $\mathbf{v}^{3D}_{1:2}$ we make the following approach
\begin{equation}
    \mathbf{v}^{3D}_{1:2} = \lambda_{w_{\perp i}h} \, h_\Omega (c_1(x_1,x_2,x_3) + c_2(x_1,x_2,x_3))\mathbf{v}^{2D}, \int_{h_1}^{h_2} c_1 = 1, \int_{h_1}^{h_2} c_2 = 0.
\label{eq:velapproach}
\end{equation}
Here, the factor $\lambda_{w_{\perp i}h}$ accounts for corrections whenever the width to height ratio is not large enough. It is introduced as a modification to take into account the paraboloid shape of the velocity field and increases in relevance as the width of the cross-section of the permeable pores decreases. For more details and the derivation of this empirical term, see \cite{Krach2025a}. The formulation used can be found in \ref{app:G}, Eq.~(\ref{eq:empirical_lambdas}).
The approach \cref{eq:velapproach} gave good results in our numerical tests, which is why we did not investigate further approaches. Generalizing to vector-valued corrections is left to future work.

The function $c_1$ is chosen as the classical parabolic profile, i.e.
\begin{equation}
    c_1(x_1,x_2,x_3) := \frac{6}{(h_2 - h_1)^3} \left( (x_3 - h_1)(h_2 - x_3) \right),
\end{equation}
with derivatives
\begin{equation}
    \frac{\partial c_1}{\partial x_3}(\mathbf{\Phi}_2) = - \frac{6}{(h_2 - h_1)^2}\, , \quad  \frac{\partial c_1}{\partial x_3}(\mathbf{\Phi}_1) =  \frac{6}{(h_2 - h_1)^2}.
\end{equation}
Using this, the drag term can be written as
\begin{equation}
    \begin{aligned}
    -\mathbf{f}^v_{drag} &= \mu \sum_{j=1}^2 (-1)^{j+1} \mathbf{F}( \nabla^{2D} h_j) \left( \frac{(-1)^{j+1}6}{(h_2 - h_1)^2} +  \frac{\partial c_2}{\partial x_3}(\mathbf{x}^{2D},h_j(\mathbf{x}^{2D}))\right) \lambda_{w_{\perp i}h} \mathbf{v}^{2D} \\
    &=\mu \sum_{j=1}^2 \mathbf{F}( \nabla^{2D} h_j) \left( \frac{6}{(h_2 - h_1)^2} +  (-1)^{j+1}\frac{\partial c_2}{\partial x_3}(\mathbf{x}^{2D},h_j(\mathbf{x}^{2D}))\right) \lambda_{w_{\perp i}h} \mathbf{v}^{2D} \\
    &= \mu \frac{6}{(h_2 - h_1)^2} \sum_{j=1}^2 \mathbf{F}( \nabla^{2D} h_j) \left( 1 + \frac{(-1)^{j+1} (h_2 - h_1)^2}{6} \frac{\partial c_2}{\partial x_3}(\mathbf{x}^{2D},h_j(\mathbf{x}^{2D}))\right) \lambda_{w_{\perp i}h} \mathbf{v}^{2D} \\
    &= \mu \frac{6}{(h_2 - h_1)^2} \sum_{j=1}^2 \mathbf{F}( \nabla^{2D} h_j) \left( 1 + \lambda_j(\mathbf{x}^{2D})\right) \lambda_{w_{\perp i}h} \mathbf{v}^{2D},
    \end{aligned}
\end{equation}
where we have introduced the function $\lambda_j$.

When assuming that precipitation happens only at the lower boundary, as discussed in \cref{ssec:experimental_setup} and assumed in this manuscript, then it holds that $h_2 = h_\Omega$, the drag term can then be written in terms of $h = h_\Omega - h_1$ as
\begin{equation}\label{eq:final_general_drag_both}
    \begin{aligned}
    -\mathbf{f}^v_{drag}
    & =  \frac{6\mu}{h(\mathbf{x})^2} \left( \left( 1 + \lambda_1(\mathbf{x}^{2D})\right) \mathbf{{F}}( \nabla^{2D} h) + \left( 1 + \lambda_2(\mathbf{x}^{2D})\right) \mathbf{I} \right) \lambda_{w_{\perp i}h}  \mathbf{v}^{2D}, \quad  \\
    -\mathbf{f}^p_{drag} & =  - \frac{\mu}{h_\Omega} \lambda^p(\mathbf{x}^{2D}) \nabla^{2D} h \, , \\
    \end{aligned}
\end{equation}
where we have used $ \nabla^{2D} h_1 = - \nabla^{2D} h$ and  $\mathbf{{F}}( -\nabla^{2D} h) = \mathbf{{F}}(\nabla^{2D} h) $, and we write $ \lambda^p :=  \lambda_1^p$, i.e. omitting the subscript since it is clear to which boundary it refers to. When assuming that $h_2 \equiv h_\Omega$ the influence of $\lambda_2$ is rather small such that we make the assumption $\lambda_2 \approx 0$. This results in
\begin{equation}\label{eq:final_drag_term}
    \begin{aligned}
    -\mathbf{f}^v_{drag}
    & = \frac{6 \mu}{h(\mathbf{x})^2} \left(  \mathbf{I} + \left( 1 + \lambda_1(\mathbf{x}^{2D})\right) \mathbf{{F}}( \nabla^{2D} h)  \right)  \lambda_{w_{\perp i}h}  \mathbf{v}^{2D}, \quad \\
    -\mathbf{f}^p_{drag} &= - \frac{\mu}{h_\Omega} \lambda^p(\mathbf{x}^{2D}) \nabla^{2D} h \, . \\
    \end{aligned}
\end{equation}

\begin{remark}
If  $\lambda_1(\mathbf{x}^{2D}) = 0$, $\lambda_{w_{\perp i}h} = 1$ then $-\mathbf{f}^v_{drag}
    = \mu \frac{6}{h^2} \left(  \mathbf{I} + \mathbf{{F}}( \nabla^{2D} h)  \right)  \mathbf{v}^{2D} $. For the case that in addition $h_1 \equiv 0 \rightarrow h(\mathbf{x}) = h_\Omega$, i.e.  $\mathbf{{F}}( \nabla^{2D} h) = \mathbf{I}$, we recover the classical drag term for a parabolic channel profile, $-\mathbf{f}^v_{drag}
    = \mu \frac{12}{h_\Omega^2}  \mathbf{v}^{2D} $.
\end{remark}

\subsection{Overview of methods and drag term formulations}
\noindent To conclude this section, we provide a summary (Table~\ref{tab:all_methods}) of methods, the governing equations, and drag terms used in the subsequent chapter for benchmarks and applications. This overview serves as a concise reference for the approaches and formulations applied throughout this study. We refer to the derived approach (Eq.~(\ref{eq:2d_brinkman_gov_equations}) with Eq.~(\ref{eq:final_drag_term})) as \STOKES throughout this manuscript. In the following sections, we also use weighted drag term approaches, as developed in \cite{Krach2025a}, in which the relevant supporting details are provided.
\begin{table}[h]
  \centering
  \footnotesize
  \begin{tabular}{ l c c c c }
    \hline
    Name & Dimension & Gov. Eq. & Drag Term ($-\mathbf{f}_{drag}$) \\ %
    poremaps & 3D & Eq.(\ref{eq:3d_gov_equations}) & $-$  \\
    OpenFOAM & 3D & Eq.(\ref{eq:3d_gov_equations}) &  $-$ \\

    Stokes, original $\mathbf{f}_{drag}$ & pseudo-3D & Eq.(\ref{eq:2d_gov_equations_with_drag}) & $ \frac{12 \mu}{h(\mathbf{x})^2} \mathbf{v}^{2D}$ \\
    Stokes, total $\mathbf{f}_{drag}$ & pseudo-3D  & Eq.(\ref{eq:2d_gov_equations_with_drag}) & $ \frac{12 \mu}{h(\mathbf{x})^2} \boldsymbol{\lambda}^s(\mathbf{x}) \mathbf{v}^{2D}$ \\
    \STOKESohne, analytical $\mathbf{f}_{drag}$ & pseudo-3D  & Eq.(\ref{eq:2d_brinkman_gov_equations}) & $ \frac{6 \mu}{h(\mathbf{x})^2} \left( \mathbf{I} + \mathbf{{F}}( \nabla^{2D} h) \right) \mathbf{v}^{2D}$ \\
    \STOKESohne, total $\mathbf{f}_{drag}$ & pseudo-3D  & Eq.(\ref{eq:2d_brinkman_gov_equations}) & $  \frac{6\mu}{h(\mathbf{x})^2} \left(  \mathbf{I} + \left( 1 + \lambda_1(\mathbf{x}^{2D})\right) \mathbf{{F}}( \nabla^{2D} h)  \right) \lambda_{w_{\perp i}h} \mathbf{v}^{2D}$  \\
    & & & $  - \frac{\mu}{h_\Omega} \lambda^p(\mathbf{x}^{2D}) \nabla^{2D} h$ \\
    \hline
  \end{tabular}

  \caption{Overview of numerical methods and drag term formulations for simulating flow in clogging porous media.
The table summarizes different approaches for fully 3D pore-scale simulations (poremaps and OpenFOAM) and pseudo-3D models using \Dumux with their governing equations, drag term formulations, and corresponding references.}
  \label{tab:all_methods}
\end{table}

\subsection{Numerical solvers}
\noindent For the computation of the permeability of the 3D domains we use OpenFOAM \cite{of2312} and poremaps \cite{krach2024a, Krach2025b}. For the \textit{Single Precipitate} benchmark (see Section~\ref{ssec:results_singleprec}) and the \textit{Segments} \textit{1} - \textit{3} (see Section~\ref{ssec:results_segments}) the permeabilities are already given in \citet{Krach2025a} and the simulation data is available in \cite{darus-4313_2024, darus-4347_2024}. For all pseudo-3D simulations we employ the open-source software \Dumuxohne, a well established Finite Volume solver which has proven its reliability in solving these problem \citep{Flemischetal2011,Kochetal2020,Weishauptetal2019}. The staggered grid \cite{harlow1965numerical} is generated using the DUNE subgrid module of the DUNE library \cite{Bastianetal2008a, Bastianetal2008b} by incorporating the images from the microfluidic experiment. Dirichlet boundary conditions are imposed for the pressure at the inlet and outlet, while no-slip conditions (zero velocity) are applied to all other boundaries, including fluid-solid interfaces.

\section{Results}\label{sec:results}
\subsection{Benchmark: Single precipitate}\label{ssec:results_singleprec}

\begin{figure}
    \centering
    \includegraphics[width=\textwidth]{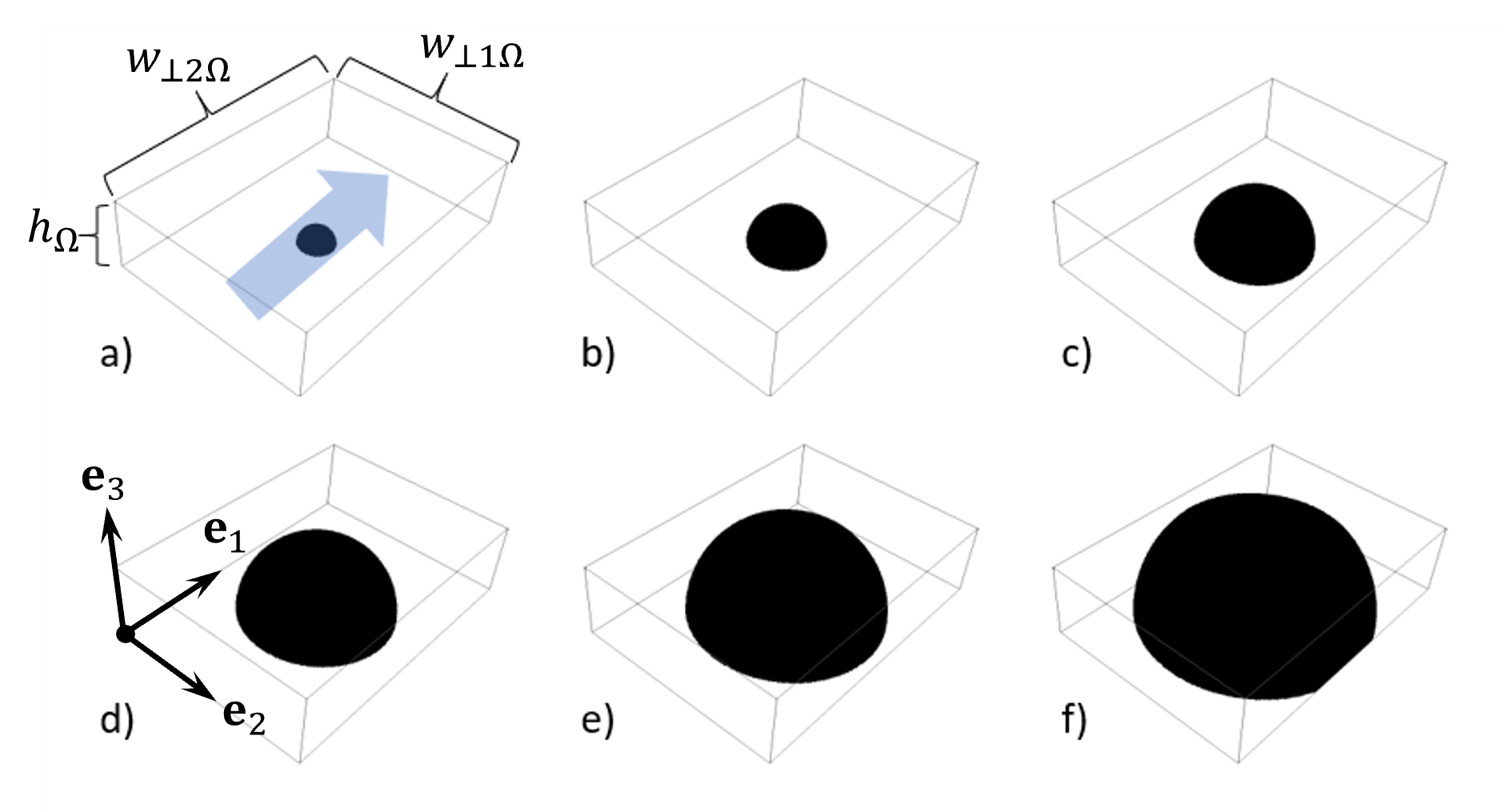}
    \caption{Dimension and geometry of the \textit{single precipitate domain}. The channel length, $w_{\perp 2 \Omega} = 140\,\mathrm{\mu m}$; height, $h_\Omega = 36\, \mathrm{\mu m}$ and width, $w_{\perp 1 \Omega} = 100\, \mathrm{\mu m}$ are fixed. a) - f) show the single precipitate with varying radii: a) $r_{prec} = h_\Omega/4$, b) $r_{prec} = h_\Omega/2$, c) $r_{prec} = 3h_\Omega/4$, d) $r_{prec} = h_\Omega$, e) $r_{prec} = 5h_\Omega/4$, f) $r_{prec} = 3h_\Omega/2$. The blue arrow indicates the flow direction.  Reprinted from \cite{Krach2025a}.}
    \label{fig::SinglePrecDomain}
\end{figure}

\noindent For the analysis of the methods, we employ the single precipitate benchmark case, as also used in \citet{Krach2025a}. We investigate the performance of different formulations, see Table~\ref{tab:all_methods}, for precipitates (semi-spheres) of varying sizes within a rectangular channel with fixed solid boundaries, see Figure \ref{fig::SinglePrecDomain}. This setup further allows us to analyze velocity profiles for quite complex cases, including low width-to-height ratios, steep gradients $\nabla^{2D} h(\mathbf{x})$, and skewed cross-sectional areas.

This configuration enables a methodical comparison of various modeling approaches and the identification of their respective limitations. The domain comprises a rectangular channel with a length $w_{\perp 2\Omega}$ of $140 \mu\mathrm{m}$, a width $w_{\perp 1\Omega}$ of $100 \mu\mathrm{m}$, and a height $h_\Omega$ of $36 \mu \mathrm{m}$. Seven distinct scenarios were examined, each incorporating a semi-spherical precipitate centered within the channel. The first scenario represents the baseline case, where the channel contains no precipitate. In the subsequent six scenarios, illustrated in Fig. \ref{fig::SinglePrecDomain}(a)–(f), the precipitate radius $r_{\text{prec}}$ increases incrementally from $h_\Omega/4$ to $3h_\Omega/2$.
The direction of flow is indicated by the blue arrow. The boundary conditions impose fixed pressures at both the inlet and outlet, thereby establishing a pressure gradient along the length of the channel. No-slip and no-penetration conditions are uniformly applied to the side, top, and bottom walls, as well as to the surface of the precipitate in all cases. The flow rate $q_1$ at the outlet is calculated, and the permeability coefficient $k_{11}$ is determined using Darcy`s law \cite{darcy1856fontaines}
\begin{equation}\label{eq:darcy}
    \mathbf{q} = - \frac{1}{\mu} \mathbf{k} \cdot \mathrm{grad} p \quad \rightarrow \quad  q_i \mathbf{e}_i = - \frac{1}{\mu} k_{ij} (\mathbf{e}_i \otimes \mathbf{e}_j) \cdot p_{,k} \mathbf{e}_k = - \frac{1}{\mu} k_{ik} p_{,k} \mathbf{e}_i \, .
\end{equation}

\begin{figure}
    \centering
    \includegraphics[width=0.98\linewidth]{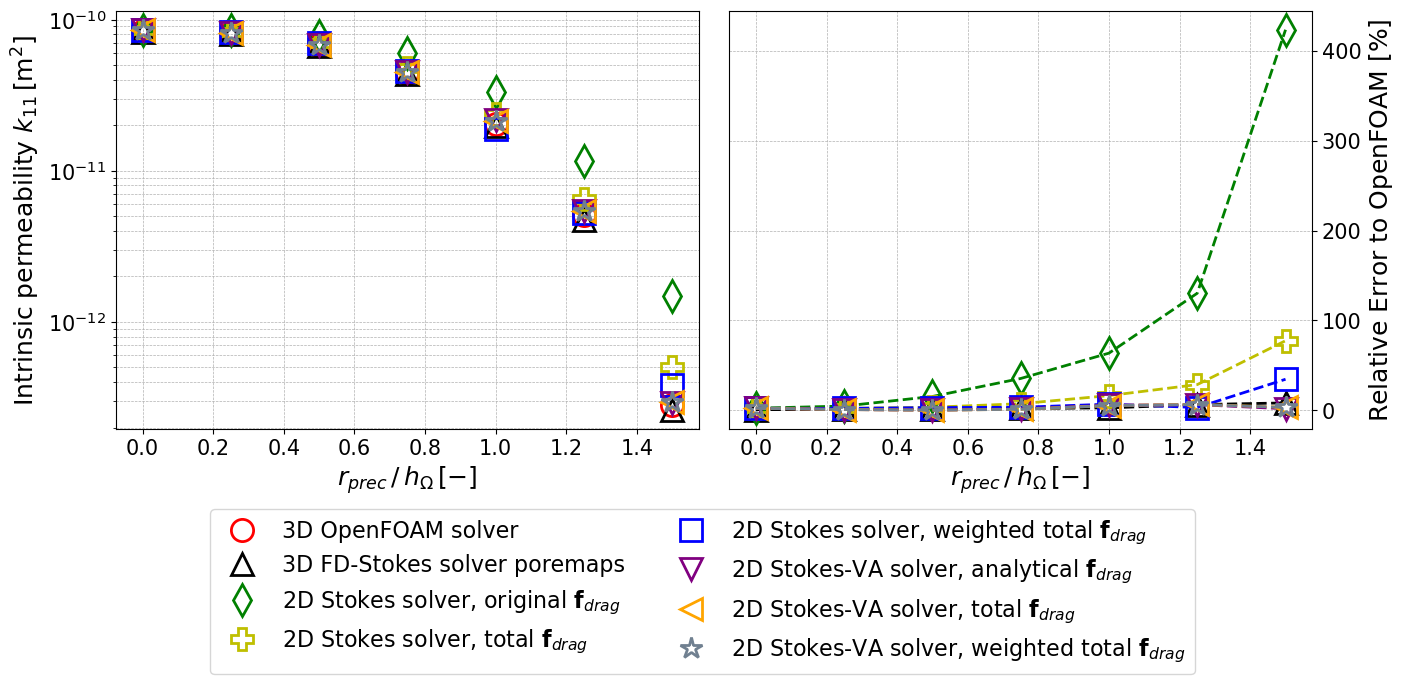}
    \caption{Permeabilities for single precipitate domain (left) and error of pseudo-3D \Dumux solvers (\STOKES and Stokes) with different $\mathbf{f}_{drag}$-formulations and 3D poremaps simulations relative to 3D OpenFOAM (right).}
    \label{fig:comparison_permeabilities}
\end{figure}

\begin{figure}
    \centering
    \includegraphics[width=0.98\linewidth]{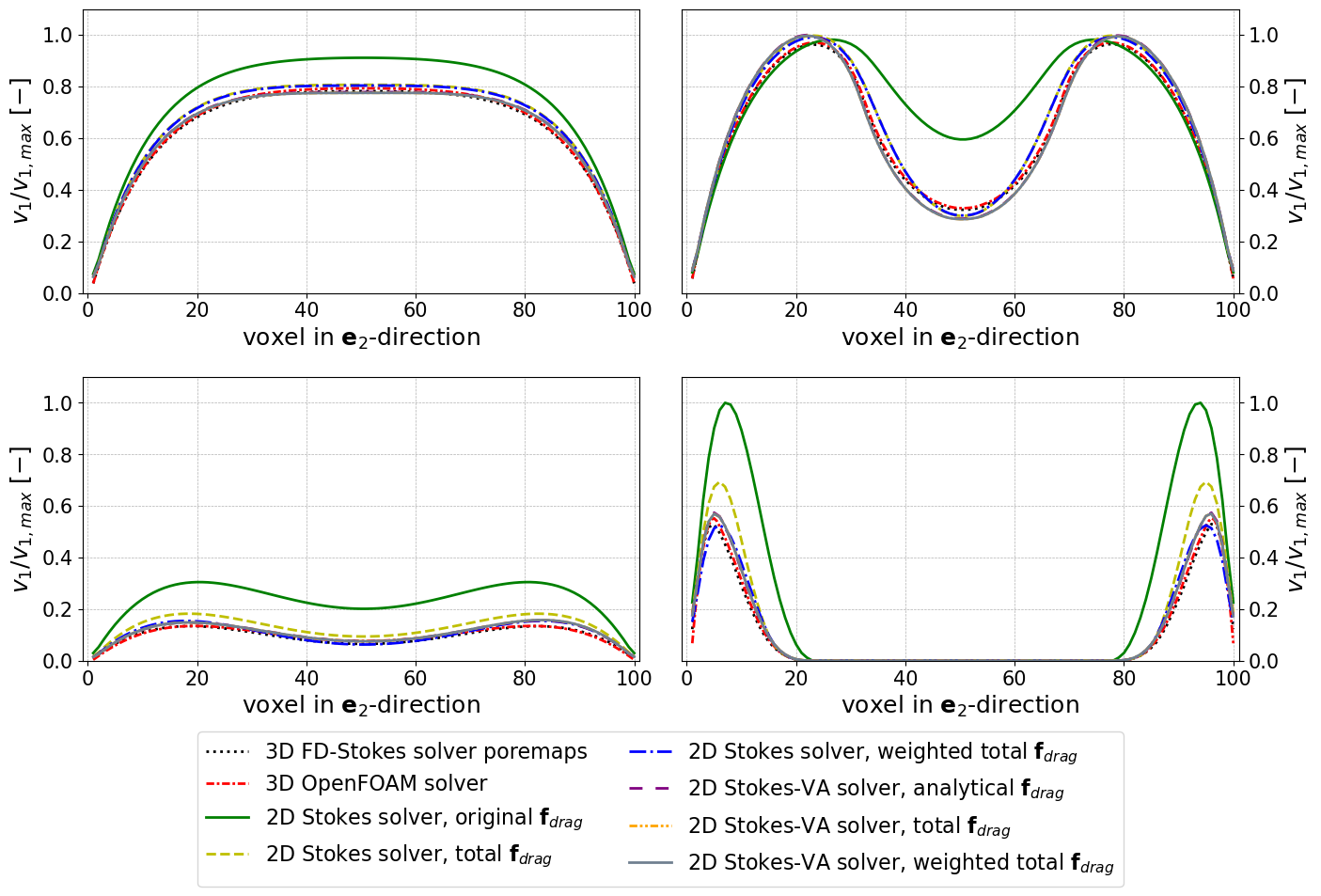}
    \caption{Normalized velocity profiles obtained using different drag formulations and solvers (\STOKES and Stokes) for inflow (left column) and the central cross-section (right column) for $r_{prec} = 1/2 \, h_\Omega$ (first row) and $r_{prec} = 5/4 \, h_\Omega$ (second row).}
    \label{fig:comparison_profiles}
\end{figure}
\noindent All values for permeabilities and relative errors are given in Table \ref{tab:permeabilities_single_prec} and Table \ref{tab:error_single_prec}, respectively. The permeability values obtained from OpenFOAM and poremaps serve as reference standards, as both methods are recognized for their accuracy in resolving flow through complex geometries. The results are compared across multiple modeling approaches, including Stokes and Stokes-VA solvers, each employing different drag force formulations.

Figure~\ref{fig:comparison_permeabilities} illustrates macroscopic results (permeability and relative deviations) while Figure~\ref{fig:comparison_profiles} presents a detailed analysis of microscopic pore fluid velocities for different cases and cross-sections allowing for a holistic analysis of the different drag formulations. For the baseline case, where no precipitate is present ($ r_{prec}=0 $), all models produce permeability values that closely align with the reference data, with only minor deviations (max. $2\%$) in permeability observed. However, as the size of the precipitate increases, the discrepancies between some models and the reference values become increasingly pronounced. The Stokes solver, when used with the original drag formulation, significantly overestimates permeability, particularly for larger precipitate sizes. At $r=3h_\Omega/2$, the error exceeds $400\%$, clearly demonstrating the inadequacy of this formulation for scenarios involving substantial flow obstructions.

The Stokes solver with the total and weighted total drag formulations, as proposed in \cite{Krach2025a} shows notable improvements in accuracy compared to the original formulation. Nevertheless, these approaches still exhibit considerable errors, especially for precipitate sizes $r \ge h_\Omega$. The errors in these cases even exceed $77\%$ and $34\%$, respectively, indicating that while these formulations represent an advancement, they are not yet sufficient for highly obstructed flows. Profiles for $r_{prec} = \frac{h_\Omega}{2}$ and $r_{prec} = \frac{5h_\Omega}{4}$ at the inflow and center cross-section are shown in Fig.~\ref{fig:comparison_profiles}. All drag formulations yield smooth profiles. Maximum velocities occur using the Stokes equations with the original drag formulation. Partial improvements appear when considering modified drag formulations. The Stokes-VA solver most accurately replicates the actual solutions obtained from the 3D solver. Only minor discrepancies are observed, specifically in the maximum values and the locations of the velocity maxima, which are slightly shifted. This is visible particularly at the domain center (Fig.~\ref{fig:comparison_profiles}, second row, right), where effective cross-sections are highly skewed.

In contrast, the Stokes-VA solver generally outperforms the Stokes solver across all tested configurations. Among the Stokes-VA solver variants, all drag formulations perform excellent with errors consistently below $7\%$ compared to the reference permeability values. The analytical drag formulation also demonstrates competitive accuracy without any $\boldsymbol{\lambda}$ corrections. Although it tends to produce slightly higher errors for some precipitate sizes, for other precipitate sizes it can be considered the most accurate pseudo-3D method. These results suggest that the \STOKES solver, independent from the adapted drag formulation used, is the most reliable approach for predicting permeability in the presence of significant flow obstructions.

In particular, the analytical approach for $\mathbf{f}_{drag}$ (see Table~\ref{tab:all_methods}, line $5$) presents a promising alternative to existing solvers and our modified methods. The Stokes-VA solver combined with this drag formulation does not rely on additional $\lambda$ terms; instead, it is fully computable from image data and completely implemented within the \Dumux framework.
As a result, it eliminates the need for preprocessing and reduces the overall computation time by approximately $50\%$ (see Section \ref{ssec:results_segments}, Table \ref{tab:time_all_methods} for exact numbers for Segments example).
Although this improvement is of minor relevance for the single precipitate benchmark, where both preprocessing (computing $\boldsymbol{\lambda}$ fields) and \Dumux simulations require only about $10$ seconds each on a state-of-the-art desktop computer, it becomes highly significant during model validation for real geometries.
Overall, the data support the conclusion that in terms of accuracy, the proposed method achieves performance comparable to that of the $\boldsymbol{\lambda}$-term variants. This suggests that the data-driven approach introduced by \citet{Krach2025a} and its reliance on extensive averaging per cross-section produces good results in terms of permeability, hence a more computationally straightforward version of the drag term is preferable. This discrepancy is evident when comparing the maximum velocities in Figure~\ref{fig:comparison_profiles}, where Stokes solvers exhibit a notable shift. Although this deviation has a negligible impact on the overall permeability, it highlights a limitation inherent to the data-driven approach that is not observed in the \STOKES solver. Apart from the proven reliability in this first benchmark, the \STOKES solver still has limitations when the domain exhibits small width to height of cross-sections. In these cases, the total drag term (including a $\lambda_w$) should always be used, as the analytical \STOKES solver does not include a compensation for this issue, resulting in exactly the same errors as given in \citet{Krach2025a}, section 3.3.1.

\begin{table}%
  \centering
  \scriptsize
  \begin{tabular}{l l | c c c c c c c }
    \hline
    solver & $\mathbf{f}_{drag} $ & $r = 0$ & $r = {h_\Omega}/{4}$ & $r = {h_\Omega}/{2}$ & $r = { 3 h_\Omega}/{4}$ & $r = h_\Omega$ & $r = {5 h_\Omega}/{4}$ & $r = {3 h_\Omega}/{2}$ \\ \hline
    OpenFOAM & - &$8.36$ & $8.09$ & $6.77$ & $4.44$ & $2.04$ & $0.507$ & $0.028$ \\
poremaps & - & $8.29$ & $8.01$ & $6.68$ & $4.34$ & $1.98$ & $0.474$ & $0.026$ \\
Stokes & original  & $8.55$ & $8.47$ & $7.81$ & $6.01$ & $3.33$ & $1.16$ & $0.148$ \\
Stokes & total  & $8.55$ & $8.26$ & $7.00$ & $4.76$ & $2.37$ & $0.652$ & $0.050$ \\
Stokes & weighted total  & $8.55$ & $8.26$ & $6.97$ & $4.59$ & $1.89$ & $0.522$ & $0.038$ \\
\STOKES & total  & $8.55$ & $8.16$ & $6.76$ & $4.49$ & $2.14$ & $0.540$ & $0.029$ \\
\STOKES & weighted total  & $8.55$ & $8.16$ & $6.76$ & $4.49$ & $2.14$ & $0.539$ & $0.029$ \\
\STOKES & analytical  & $8.55$ & $8.16$ & $6.78$ & $4.51$ & $2.15$ & $0.539$ & $0.029$ \\
  \end{tabular}
  \caption{Permeabilities for the single precipitate domains with varying radii of the precipitate in $\left[ 10^{-11} \, m^2 \right]$.}
  \label{tab:permeabilities_single_prec}
\end{table}

\begin{table}%
  \centering
  \scriptsize

  \begin{tabular}{l l | c c c c c c c }
    \hline
    solver & $\mathbf{f}_{drag}$ & $r = 0$ & $r = {h_\Omega}/{4}$ & $r = {h_\Omega}/{2}$ & $r = { 3 h_\Omega}/{4}$ & $r = h_\Omega$ & $r = {5 h_\Omega}/{4}$ & $r = {3 h_\Omega}/{2}$ \\ \hline
poremaps & - & $0.79$ & $0.97$ & $1.29$ & $2.07$ & $2.75$ & $6.37$ & $8.27$ \\
Stokes & original  & $2.37$ & $4.76$ & $15.41$ & $35.45$ & $63.65$ & $130.30$ & $423.42$ \\
Stokes & total  & $2.31$ & $2.16$ & $3.46$ & $7.40$ & $16.41$ & $28.66$ & $77.46$ \\
Stokes & weighted total  & $2.31$ & $2.14$ & $3.07$ & $3.41$ & $6.87$ & $3.15$ & $34.51$ \\
\STOKES & total  & $2.34$ & $0.87$ & $0.15$ & $1.19$ & $5.30$ & $6.63$ & $3.39$ \\
\STOKES & weighted total  & $2.34$ & $0.87$ & $0.15$ & $1.19$ & $5.24$ & $6.84$ & $3.24$ \\
\STOKES & analytical  & $2.34$ & $0.93$ & $0.16$ & $1.60$ & $5.69$ & $6.35$ & $1.75$ \\
  \end{tabular}
  \caption{Errors relative to the OpenFOAM reference for the single precipitate domains with varying radii of the precipitate in $[\%]$.}
  \label{tab:error_single_prec}
\end{table}

\subsection{Validation: Segments}\label{ssec:results_segments}
\begin{figure}[h!]
    \centering
    \includegraphics[width=0.9\linewidth]{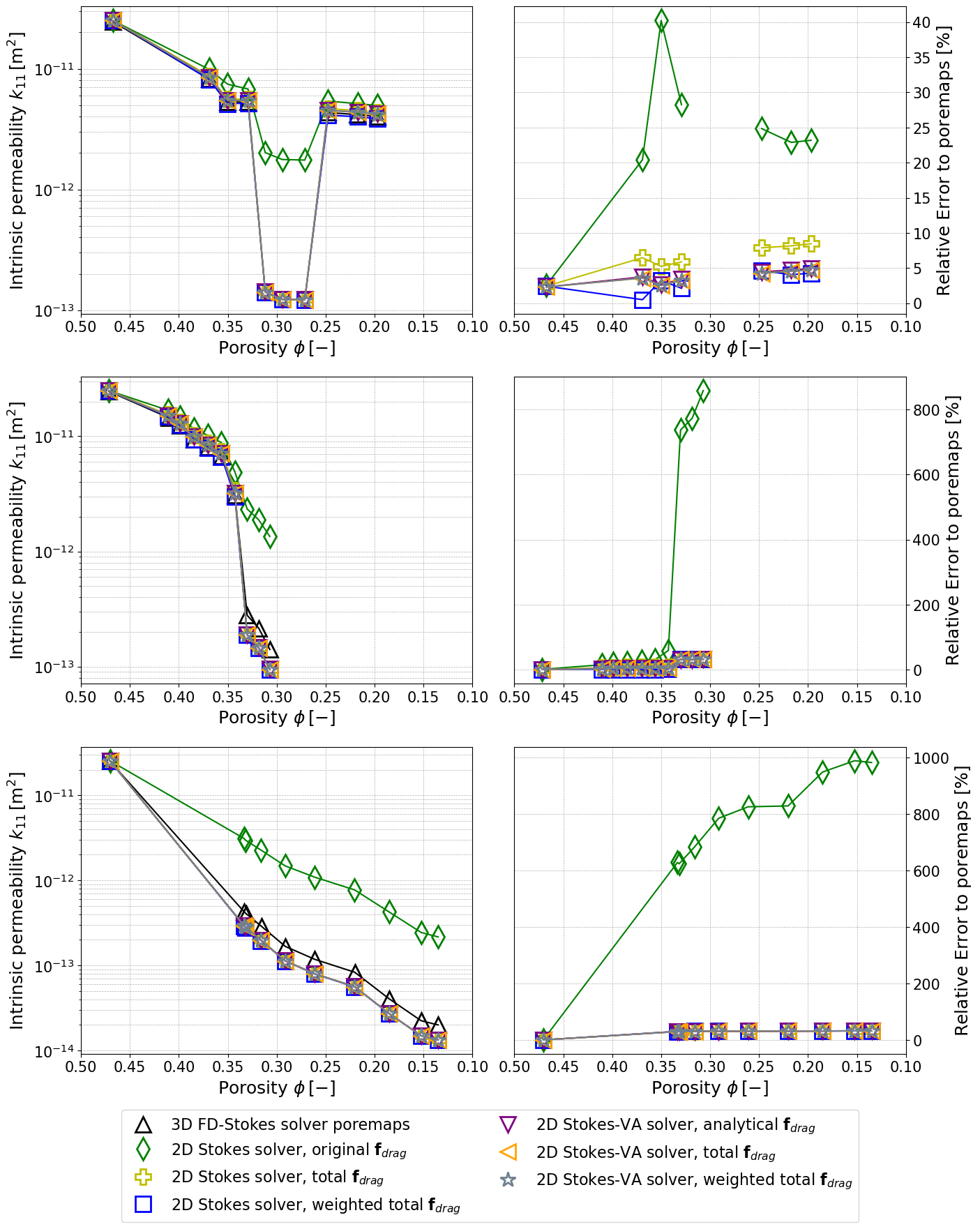}
    \caption{The relationship between porosity and permeability is examined across three distinct segments during the clogging process.
As time progresses, porosity $\phi$ consistently declines, which is shown in the $\phi - k_{11}$ relationships on the left side. The right side presents the relative error of pseudo-3D simulations conducted in \Dumuxohne , employing various drag formulations $\mathbf{f}_{drag}$ and different governing equations, in comparison to fully resolved 3D poremaps simulations.}
    \label{fig:results_segments}
\end{figure}

The evolution of permeability over the course of the experiment, corresponding to decreasing porosity, and the associated relative errors for the \textit{Segments 1-3} (geometry in Figure~\ref{fig::cellsegments_experiment}) are presented in Figure~\ref{fig:results_segments}. Additional information on the experimental setup and results are provided in Section \ref{ssec:experimental_setup} and in greater detail in \citet{weinhardt2022spatiotemporal}.
\textit{Segment 1} is characterized by a heterogeneous precipitation pattern and the formation of a preferential flow path. There is negligible flow and intense precipitation in the bottom part in the inflow region and flow is dominated by the preferential flow path that can be interpreted as an irregular open rectangular channel. \textit{Segment 2} and \textit{Segment 3} show a homogeneous precipitation distribution, with \textit{Segment 2} characterized by lower precipitation and terminal porosity due to its remote location, while \textit{Segment 3}, like \textit{Segment 1}, is located in the front/inflow region and is experiencing high precipitation.

\paragraph*{Permeability} For this analysis, only the $k_{11}$ component of the permeability tensor is computed, which is sufficient to validate the applicability of the proposed approach. The results for different formulations of drag terms are presented, calculated for $10$ snapshots of each segment throughout the experiment. The reference values are computed by poremaps and taken from \citet{darus-4347_2024,darus-4313_2024}. Permeability values are closely clustered, indicating that all variants that do not use the original version of the drag term (green diamonds) deliver good results across the entire porosity range.

\begin{table}[h]
  \centering
  \scriptsize
\begin{tabular}{l l | c c c c c c c c c c}
\hline
Solver & $\mathbf{f}_{drag}$ & \multicolumn{10}{c}{Snapshots of \textit{Segment 1}: relative errors to 3D solver $[-]$} \\
\hline
Stokes & original & $2.56$ & $20.41$ &  $40.23$ & $28.25$ & $-$ & $-$ & $-$ & $24.85$ &  $22.89$ & $23.19$ \\
Stokes & total & $2.43$ & $6.48$& $5.24$& $5.89$& $-$ & $-$ & $-$ & $7.88$  & $8.17$ & $8.47$ \\
Stokes & weighted total & $2.43$ &  $0.51$ & $3.24$ &  $2.13$ & $-$ & $-$ & $-$ & $4.64$ & $4.07$ & $4.28$ \\
\STOKESohne & analytical  & $2.31$ & $3.76$ & $2.70$ & $3.41$ & $-$ & $-$ & $-$ & $4.47$ & $4.73$ & $ 4.91$ \\
\STOKESohne & total & $2.31$ & $3.60$ & $2.53$ & $3.28$ & $-$ & $-$ & $-$ & $4.29$ & $4.57$ & $4.72$ \\
 \STOKESohne & weighted total & $2.31$ & $3.59$ & $2.52$ & $3.28$ & $-$ & $-$ & $-$ & $4.28$ & $ 4.57$ & $4.72$ \\
\hline

 & &  \multicolumn{10}{c}{Snapshots of \textit{Segment 2}: relative errors to 3D solver $[-]$} \\
\hline
 Stokes & original  & $1.72$ & $15.65$ & $19.19$ & $23.03$ & $27.06$ & $31.06$ & $59.57$ & $736.96$ & $772.95$ & $857.97$ \\
Stokes &  total & $1.58$ & $6.25$ & $7.58$ & $8.37$ & $9.06$ & $10.14$ & $10.05$ & $32.14$ & $32.14$ & $32.64$ \\
Stokes & weighted total & $1.58$ & $1.87$ & $1.40$ & $0.24$ & $0.86$ & $0.88$ & $2.48$ & $32.54$ & $32.54$ & $32.88$ \\
 \STOKESohne & analytical & $1.47$ & $3.81$ & $4.42$ & $4.82$ & $5.08$ & $5.54$ & $4.89$ & $31.84$ & $31.87$ & $32.44$ \\
\STOKESohne & total & $1.47$ & $3.65$ & $4.24$ & $4.61$ & $4.85$ & $5.24$ & $4.50$ & $32.25$ & $32.57$ & $33.35$ \\
\STOKESohne & weighted total & $1.47$ & $3.65$ & $4.24$ & $4.61$ & $4.84$ & $5.23$ & $4.50$ & $32.25$ & $32.57$ & $33.35$ \\
\hline
&  & \multicolumn{10}{c}{Snapshots of \textit{Segment 3}: relative errors to 3D solver $[-]$} \\
\hline
 Stokes & original  & $1.66$ & $629.25$ & $624.62$ & $684.55$ & $785.49$ & $826.14$ & $829.14$ & $949.12$ & $989.18$ & $982.67$ \\
Stokes & total  & $1.54$ & $31.20$ & $30.93$ & $31.65$ & $32.10$ & $32.37$ & $32.24$ & $33.01$ & $33.47$ & $33.32$ \\
 Stokes & weighted total  & $1.54$ & $31.79$ & $31.53$ & $32.21$ & $32.56$ & $32.68$ & $32.66$ & $33.18$ & $33.52$ & $33.36$ \\
 \STOKESohne & analytical  & $1.42$ & $30.85$ & $30.53$ & $31.43$ & $31.95$ & $32.19$ & $32.09$ & $32.81$ & $33.34$ & $33.24$ \\
 \STOKESohne & total  & $1.42$ & $31.08$ & $30.75$ & $31.69$ & $32.24$ & $32.35$ & $32.24$ & $32.95$ & $33.43$ & $33.30$ \\
\STOKESohne & weighted total  & $1.42$ & $31.08$ & $30.75$ & $31.69$ & $32.24$ & $32.35$ & $32.25$ & $32.95$ & $33.43$ & $33.30$ \\
\hline
\end{tabular}
\label{tab:errors_segments}
\caption{Errors relative to the poremaps reference for all $10$ snapshots the \textit{Segments 1-3} in $[\%]$ for different \Dumuxohne~solvers and drag terms.}
\end{table}
Note, that the consistent error of $\approx 30 \%$ is due to the inaccuracies of 3D results for cases with insufficient resolution in some parts of the domain. An analysis of the error is provided in \citet{Krach2025b}.
For \textit{Segment 1}, data are missing, in the range of $0.32 < \phi < 0.25$, for relative errors, since the 3D solver did not converge for these nearly fully blocked domains. In the experiment, a junk of precipitate blocks the preferential flow path for a short amount of time, which eventually is flushed out and the path is cleared.  Relative errors are reduced from $\approx 25\%$ to $\approx 5\%$ for Stokes-VA solver with the analytical $\mathbf{f}_{drag}$. Errors are generally smaller for \textit{Segment 1} since it is dominated by the preferential flow path, whose geometry does not severely violate the constraints of the original formulation $\mathbf{f}_{drag}$. For \textit{Segments 2,3} all adapted methods show immense improvements compared to the original drag term as soon as precipitation has progressed to such an extent that it is blocking pore throats. This is the case for approximately $\phi < 0.35$. Once this threshold is reached, there are significant discrepancies between the original $\mathbf{f}_{drag}$ (up to approximately $1000\%$ deviation) and all adapted versions (maximum $33\%$ error), which, however, hardly differ from one another (differences in the range of $\pm 3\%$). Given the negligible discrepancies among the evaluated methods, the corresponding curves in Figure~\ref{fig:results_segments} are nearly indistinguishable. For clarity, the relative deviations between the methods and the 3D solver are summarized in Table~\ref{tab:errors_segments}.

\paragraph*{Computational efficiency}
Table \ref{tab:time_all_methods} presents a comparative overview of average computation times for different formulations of drag terms and solvers, including preprocessing time (with the python module \texttt{localdrag} \cite{darus-4313_2024}). Among pseudo-3D methods, the analytical \STOKES approach has a clear advantage: it achieves a total computation time of $13.5$ minutes, while completely eliminating the need for the preprocessing. In contrast, alternative methods such as all Stokes and all remaining \STOKES models that use adapted (including empirical $\boldsymbol{\lambda}$) drag terms require additional preprocessing steps ($11.0 - 11.5$ minutes). This is resulting in overall computation times of $24.5 – 25.5$ minutes, nearly $50 \%$ more than the analytical \STOKES method.
This efficiency gain is of particular significance for large-scale simulation campaigns that involve hundreds or even thousands of individual simulations. Beyond raw computation time, elimination of preprocessing simplifies the organizational workflow as it removes the need for additional data preparation steps. This reduces the risk of errors associated with preprocessing pipelines, and streamlines the automation of simulation campaigns. It is especially valuable in iterative optimization processes or parameter studies, where rapid turnaround times are essential.

The analytical \STOKES method maintains solver performance comparable to other approaches, ensuring that the reduction in computational effort does not come at the expense of accuracy, which makes it an ideal choice for most applications (particularly \textit{Segments 1-3}). The poremaps data, computed on a high-performance cluster, are not directly comparable due to differences in hardware but serve as a benchmark for specialized, resource-intensive applications.
Based on these results it is suggested to employ the analytical \STOKES method, as it offers the best combination of computational efficiency, accuracy, and operational simplicity, making it highly suitable for large-scale simulation campaigns.

\begin{table}[]
  \centering
  \footnotesize
  \begin{tabular}{ l c c c c }
    \hline
    Name & Dimension & Preprocessing & Solver & Overall \\ \hline
    poremaps & 3D & - & $540$  & $540$ \\
    \Dumux Stokes, original $\mathbf{f}_{drag}$ & pseudo-3D & - & $13.0$  & $13.0$  \\
    \Dumux Stokes, total $\mathbf{f}_{drag}$ & pseudo-3D  & $11.0$ & $13.5$  & $24.5$  \\
    \Dumux Stokes, weighted total $\mathbf{f}_{drag}$ & pseudo-3D  & $11.5$ & $13.5$  & $25.0$ \\
    \Dumux \STOKESohne, analytical $\mathbf{f}_{drag}$ & pseudo-3D & - & $13.5$  & $13.5$  \\
    \Dumux \STOKESohne, total $\mathbf{f}_{drag}$ & pseudo-3D  & $11.5$ & $13.5$  & $25.0$ \\
    \Dumux \STOKESohne, weighted total $\mathbf{f}_{drag}$ & pseudo-3D  & $11.0$ & $14.5$ & $25.5$ \\

    \hline
  \end{tabular}

  \caption{Overall average computation times, including preprocessing, for different drag terms and solvers/methods. The poremaps data is not directly comparable to the other results, as they were computed on a high-performance cluster (experimental compute cluster of the SimTech Cluster of Excellence (University of Stuttgart) on a fully allocated node with $128$ cores ($2 \times 64$ cores, AMD EPYC 7702)) rather than a state-of-the-art desktop computer (single core - AMD RYZEN9 3900x) that was used for all {\Dumux} simulations. }
  \label{tab:time_all_methods}
\end{table}

\subsection{Application: Intrinsic permeability of entire domain }\label{ssec:results_wholedomain}
This section serves to demonstrate the capabilities of the \STOKES solver using the analytical \fdrag term. We simulate the entire experimental domain, Fig.~(\ref{fig::cellsegments_experiment}) for a total of nine snapshots. Experimental results, snapshot data, and additional information are available in \cite{darus-1799_2022}.
As described in Section~\ref{ssec:experimental_setup}, when preparing the computational domain from the 2D projections of the experimental images, we assume that the precipitates never fully span the cell height. However, the exact height of the precipitates cannot directly be determined from the images only. Therefore during image processing, a maximum precipitate height must be defined.
Based on preliminary tests in which the maximum height was varied, we found that a value of 85\% of the total cell height is suitable.
In this example, we therefore prescribe a maximum precipitate height of 85\% of the cell height, which yields a minimum aperture of \( 5.25\,\mu\mathrm{m} \) throughout the domain. This parameter could, in principle, be calibrated systematically using experimental permeability data; however, such an analysis lies beyond the scope of the present work and will be addressed in future studies.

\begin{figure}
    \centering
    \includegraphics[width=0.9\linewidth]{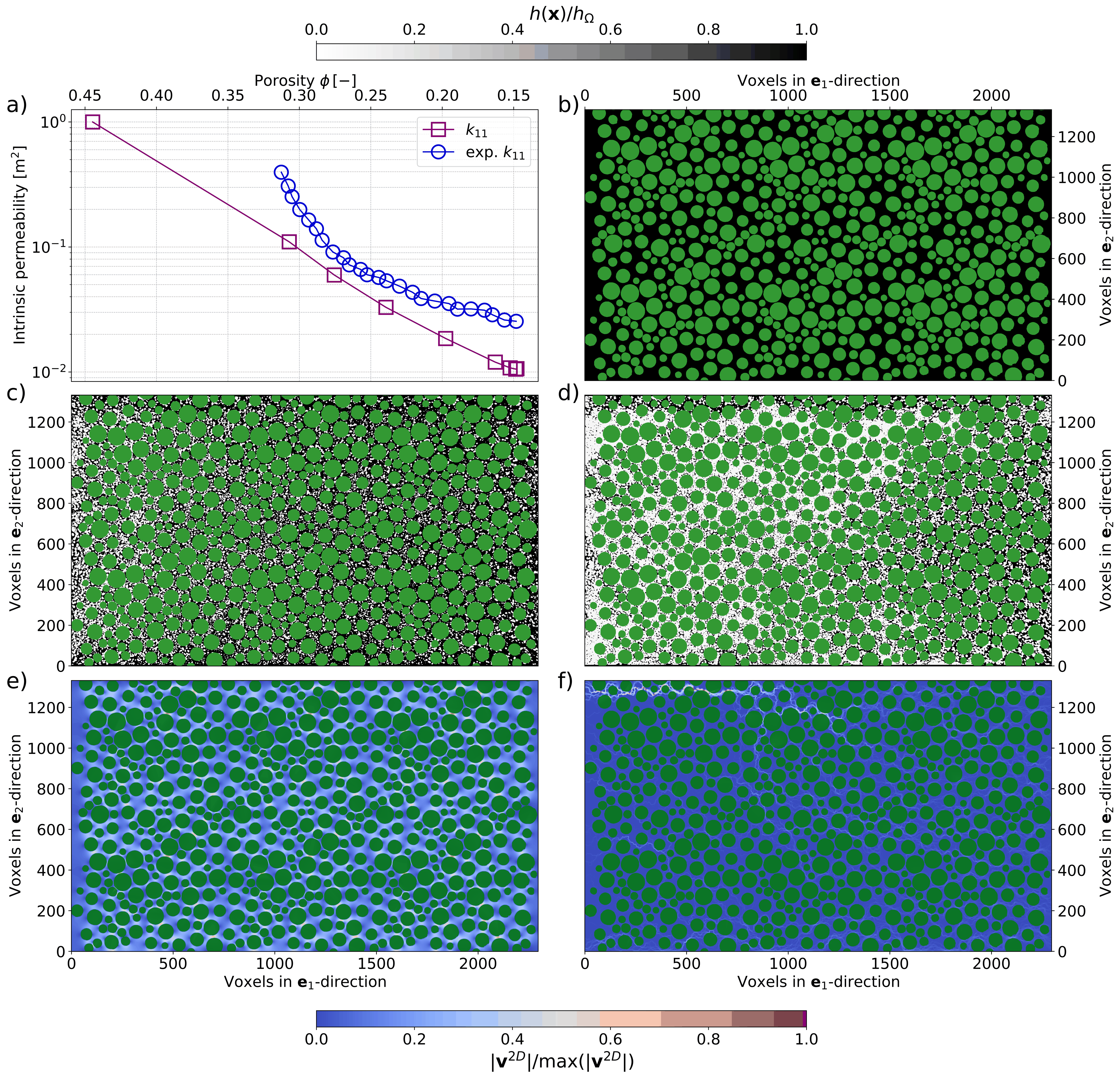}
    \caption{Entry $k_{11}$ of the intrinsic permeability tensor (a) throughout the experiment with decreasing porosity. Snapshots of the domain at different timestamps (b$-$d), whereby normalized height $h(\mathbf{x})/h_\Omega$ is given in gray values and the solid columns representing the porous material are green. Normalized magnitude of the fluid velocity $\mathbf{v}^{2D}$ at $\phi = 0.44$ (e)  and $\phi = 0.28$ (f) as a result of a pressure gradient in $\mathbf{e}_1$-direction. }
    \label{fig:results_wholedomain}
\end{figure}

The average computation time in \Dumux required per simulation is only $250$ seconds when executed on a $40$-core node, allowing efficient computation of permeability for the entire microfluidic domain, which comprises $2296 \times 1333$ voxels with a voxel edge length of 9\(\mu m\). This approach facilitates the analysis of a sweep of snapshots captured during the experiment. Snapshots of the domain are presented in Figure~\ref{fig:results_wholedomain} b) $-$ d) to illustrate the precipitation pattern (white). The normalized magnitude of the fluid velocity $\mathbf{v}^{2D}$ for snapshots with $\phi = 0.44$ (e) and $\phi = 0.28$ (f) is given in the bottom row. These visualize the transition of the flow pattern from a broadly fanned out flow, spanning the total extend of the domain, to an almost completely blocked state in the front part, except for a preferential flow pathway. At this state of the experiment the entire flow has to pass through this bottleneck of the domain, located near the upper boundary. When accounting for the uncertainties and unknowns associated with the domain and assumptions of the governing equations, as well as the errors introduced by discretization and the experimental setup, we conclude that the experimental and simulation data in Figure~\ref{fig:results_wholedomain} a) are in good agreement. However, deviations may imply that the maximum height of the precipitation may not be as uniform as assumed (consistently 85$\%$) in our model.
As mentioned above, a more detailed analysis and calibration of the maximal height of the precipitates lies beyond the scope of this study.

These findings, enabled by a highly efficient \STOKES solver and drag term, allow a fast and comprehensive assessment of the dependence of permeability on heterogeneous precipitation patterns.

\section{Conclusion and outlook}
In this paper a rigorous mathematical vertical averaging procedure is presented that substantially improves pseudo-3D simulations of clogging porous media in terms of accuracy and computation time. Starting from the fully 3D Stokes equations, we derive vertically-averaged governing equations and vertically-averaged drag terms that account for the local height of the domain $h(\mathbf{x})$, as well as its gradients $\mathrm{grad}\, h(\mathbf{x})$. The presented approach, based on an analytically derived drag term, can be further enhanced by incorporating the data-driven methodology. Even without data-driven extension, the analytical drag term demonstrates robust performance in the benchmarks presented (\textit{Single Precipitate}, Section~\ref{ssec:results_singleprec}) and application cases (\textit{Segments 1 to 3}, Section~\ref{ssec:results_segments}), maintaining deviations below $7\%$ for permeabilities. In terms of computational efficiency, this method reduces the computational cost by approximately $50\%$ compared to other pseudo-3D solvers with similar accuracy and by $97\%$ relative to the 3D solver poremaps that deliver similar accuracy. The method is considered ideal for presented applications since it combines the efficiency of a 2D solver with the precision of a 3D solver and can be applied to arbitrarily complex geometries.
By applying the presented framework to a full experimental domain, we, with minimal computational resource expenditure, shine light on the impact of local precipitation variability on permeability. As a next step, the framework could be applied to further experiments of this kind also including simulations with rotated pressure gradients to compute the entire second order permeability tensor or supplemented with a transport model, thereby further deepening our understanding of EICP in porous materials.

\newpage

\appendix
\section{Notations}
We introduce the following parameterizations for the lower and upper boundaries.
\begin{equation}
\mathbf{\Phi}_j(\mathbf{x}^{2D}) := (\mathbf{x}^{2D},h_j(\mathbf{x}^{2D})) \, .
    \label{eq:boundaryparametrization}
\end{equation}
With $f(\mathbf{\Phi}_j) := f \circ\mathbf{\Phi}_j$ we define the concatenation of some function $f$ with the parameterization $\mathbf{\Phi}_j$.

\section{Leibniz rule for divergence}
Using the Leibniz rule for parameter integrals and the no-slip velocity condition gives
\begin{equation}
\begin{aligned}
\int_{h_1}^{h_2} \nabla^{2D} \cdot \mathbf{v}_{1:2}^{3D} \, \mathrm{d}x_3 &= \sum_{j=1}^2 \frac{\mathrm{d}}{\mathrm{d} x_j}  \int_{h_1}^{h_2} v_j^{3D} \, \mathrm{d}x_3 - \left(\frac{\partial h_2}{\partial x_j} v_j^{3D}(\mathbf{\Phi}_2)  - \frac{\partial h_1}{\partial x_j} v_j^{3D}(\mathbf{\Phi}_1)\right) \\
&=  \sum_{j=1}^2 \frac{\mathrm{d}}{\mathrm{d} x_j}  \int_{h_1}^{h_2} v_j^{3D} \, \mathrm{d}x_3 \, = h_\Omega \nabla^{2D} \cdot  \mathbf{v}^{2D} .
\end{aligned}
\label{eq:div-Leibniz}
\end{equation}

\section{Leibniz rule for Laplacian}
Using the Leibniz rule for parameter integrals gives
\begin{equation}
\begin{aligned}
\int_{h_1}^{h_2} \Delta^{2D} \, v_i^{3D} \, \mathrm{d}x_3 = \sum_{j=1}^2 \frac{\mathrm{d}}{\mathrm{d} x_j}  \int_{h_1}^{h_2} \frac{\partial v_i^{3D}}{\partial x_j} \, \mathrm{d}x_3 - \left(\frac{\partial h_2}{\partial x_j}\frac{\partial v_i^{3D}}{\partial x_j}(\mathbf{\Phi}_2)  - \frac{\partial h_1}{\partial x_j}\frac{\partial v_i^{3D}}{\partial x_j}(\mathbf{\Phi}_1)\right) \, .
\end{aligned}
\end{equation}
Using again the Leibniz rule together with the no-slip boundary condition gives
\begin{equation}
\label{eq:leibnizlaplace}
\begin{aligned}
&\int_{h_1}^{h_2} \Delta^{2D} \, v_i^{3D} \, \mathrm{d}x_3 \\
    &= \sum_{j=1}^2 \frac{\mathrm{d}^2}{\mathrm{d} x^2_j}  ( h_{\Omega} v_i^{2D} ) - \left(\frac{\partial h_2}{\partial x_j}\frac{\partial v_i^{3D}}{\partial x_j} (\mathbf{\Phi}_2) - \frac{\partial h_1}{\partial x_j}\frac{\partial v_i^{3D}}{\partial x_j}(\mathbf{\Phi}_1)\right) \, .
\end{aligned}
\end{equation}

\section{Leibniz rule for pressure term}
Applying the Leibniz rule for the pressure term gives
\begin{equation}
\begin{aligned}
\int_{h_1}^{h_2} \frac{\partial}{\partial x_i} p^{3D} \, \mathrm{d}x_3 &=
\frac{\mathrm{d}}{\mathrm{d} x_i}\int_{h_1}^{h_2} p^{3D} \, \mathrm{d}x_3  - \left(\frac{\partial h_2}{\partial x_i} p^{3D}(\mathbf{\Phi}_2) - \frac{\partial h_1}{\partial x_i} p^{3D}(\mathbf{\Phi}_1) \right) \\
&= \frac{\mathrm{d}}{\mathrm{d} x_i} \left( (h_2-h_1) p^{2D}  \right) - \left(\frac{\partial h_2}{\partial x_i} p^{3D}(\mathbf{\Phi}_2) - \frac{\partial h_1}{\partial x_i} p^{3D}(\mathbf{\Phi}_1) \right) \\
&= (h_2-h_1) \frac{\partial }{\partial x_i} p^{2D} + \left( \frac{\partial h_2}{\partial x_i} \big(p^{2D} - p^{3D}(\mathbf{\Phi}_2)\big) - \frac{\partial h_1}{\partial x_i}\big(p^{2D} - p^{3D}(\mathbf{\Phi}_1)\big) \right) \, .
\end{aligned}
\label{eq:leibnizpressure}
\end{equation}

\section{Velocity derivatives at boundaries}
Due to the no-slip condition we have
\begin{equation}
    v^{3D}_i(x_1,x_2,h_j(x_1,x_2)) \equiv 0 \Rightarrow \frac{\mathrm{d}}{\mathrm{d} x_k} v^{3D}_i(x_1,x_2,h_j(x_1,x_2)) \equiv 0, \quad \forall \, i,j,k \in \lbrace 1,2 \rbrace.
\end{equation}
The total derivative can be rewritten as
\begin{equation}
\label{eq:velderivb}
\begin{aligned}
    0 &\equiv \frac{\mathrm{d}}{\mathrm{d} x_k} v^{3D}_i(x_1,x_2,h_j(x_1,x_2)) = \frac{\partial v^{3D}_i}{\partial x_k}(\mathbf{\Phi}_j) + \frac{\partial v^{3D}_i}{\partial x_3}(\mathbf{\Phi}_j) \frac{\partial h_j}{\partial x_k} \\
    &\Rightarrow \frac{\partial v^{3D}_i}{\partial x_k}(\mathbf{\Phi}_j) = - \frac{\partial v^{3D}_i}{\partial x_3}(\mathbf{\Phi}_j) \frac{\partial h_j}{\partial x_k}   \quad \forall  \,  i,j,k \in \lbrace 1,2 \rbrace.
\end{aligned}
\end{equation}

\section{Pressure at boundaries}
The momentum balance equation for $v^{3D}_3$ is given as
\begin{equation}
    \mu \Delta^{3D} v_3^{3D} =  \frac{\partial}{\partial x_3} p^{3D}.
\end{equation}
Using the splitting $p^{3D} = p^{3D,1} + \mu \lambda^p$ such that
\begin{equation}
    \mu \frac{\partial^2}{\partial x_3^2} v_3^{3D} =  \frac{\partial}{\partial x_3} p^{3D,1}, \quad  \Delta^{2D} v_3^{3D} =  \frac{\partial}{\partial x_3} \lambda^p.
\end{equation}
This allows to calculate
\begin{equation}
    p^{3D,1} = \mu \frac{\partial}{\partial x_3} v_3^{3D} + c(x_1,x_2) \, .
\end{equation}
The function $c(x_1,x_2)$ is chosen such that the vertically averaged pressure is given as $p^{2D}$ and thus
\begin{equation}
    p^{3D,1} = \mu \frac{\partial}{\partial x_3} v_3^{3D} + p^{2D}(x_1,x_2).
\end{equation}
Using the continuity equation allows to replace $\frac{\partial}{\partial x_3} v_3^{3D}$ such that
\begin{equation}
    p^{3D,1} =  p^{2D}(x_1,x_2) - \mu ( \frac{\partial}{\partial x_1} v_1^{3D} + \frac{\partial}{\partial x_2} v_2^{3D}).
\end{equation}
The whole solution is therefore given as
\begin{equation}
    p^{3D} =  p^{3D,1} + \mu\lambda^p =  p^{2D}(x_1,x_2) - \mu ( \frac{\partial}{\partial x_1} v_1^{3D} + \frac{\partial}{\partial x_2} v_2^{3D}) + \mu\lambda^p,
\end{equation}
with $\int_{h_1}^{h_2} \lambda^p = 0$.

At the boundaries using \cref{eq:velderivb} and the parameterizations \cref{eq:boundaryparametrization} gives
\begin{equation}
\label{eq:pressurebc}
\begin{aligned}
    p^{3D}(\mathbf{\Phi}_j) &=  p^{2D}  + \mu \left( \frac{\partial v^{3D}_1}{\partial x_3} (\mathbf{\Phi}_j) \frac{\partial h_j}{\partial x_1}  + \frac{\partial v^{3D}_2}{\partial x_3}(\mathbf{\Phi}_j) \frac{\partial h_j}{\partial x_2}\right) +  \mu\lambda^p(\mathbf{\Phi}_j)\\
    &= p^{2D}+ \mu \frac{\partial }{\partial x_3}\mathbf{v}^{3D}_{1:2}(\mathbf{\Phi}_j)\cdot \nabla^{2D}h_j +  \mu\lambda_j^p,
\end{aligned}
\end{equation}
where we have used $\lambda_j^p := \lambda^p(\mathbf{\Phi}_j)$.

In most cases $\frac{\partial^2}{\partial x_i^2} v^{3D}_3 << \frac{\partial^2}{\partial x_3^2} v^{3D}_3$, $i=1,2$, such that $\lambda_j^p \approx 0$ is a good approximation.

\section{Empirical relations}\label{app:G}
\noindent Despite the improved vertically averaged derivation for the drag terms and governing equations, there are 3D aspects in fluid flow that cannot be fully captured by our improved pseudo-3D approach. While these aspects are significantly less pronounced for the models including drag terms described above (Eq.~\ref{eq:final_drag_term}), they remain present. Although we recommend using the so called analytical \STOKES formulation without any $\lambda$ correction factors, data-informed correction may still be relevant for specific applications. This is the case for e.g. the \textit{Single Precipitate} case in Sec.~\ref{ssec:results_singleprec}, where this improves velocity fields for large $r_{prec}$. Following the same procedure and benchmark cases as \citet{Krach2025a}, we derive the following empirical relationships for the $\lambda_1$ and $\lambda_w$:

\begin{equation}\label{eq:empirical_lambdas}
\begin{aligned}
    \lambda_{w_{\perp i}h} & =  \frac{1.2}{\mathrm{exp}(2.25 \, w_{\perp i \Omega}/h_\Omega)} + 1 \, , \\
    \lambda_{1,i} & =  0.13 \, \nabla_i^{2D} \left(h(\mathbf{x})\right)  + 1 \, ,\\
    \lambda^p_i & = 0.1 \, \mathrm{exp}\left( 0.49 \,  \nabla_i^{2D}\left( h(\mathbf{x})\right)\right) \, ,  \quad  \text{where} \,i \in \{1,2\}
\end{aligned}
\end{equation}

\section*{Data availability}
The complete source code including input data of the shown application examples that support the findings of this study is openly available in the Data Repository of the University of Stuttgart (DaRUS) at \url{https://doi.org/10.18419/DARUS-5624} \citep{Krach2026b} and \url{https://doi.org/10.18419/DARUS-5626} \citep{Krach2026a}.

\section*{Author Contributions}
D. Krach: Conceptualization, Methodology, Investigation, Software, Validation, Data Curation, Writing - Original Draft, Visualization, Project administration. F. Weinhardt: Conceptualization, Software, Data Curation, Writing - Original Draft. H. Class: Conceptualization, Writing - Review \& Editing, Supervision, Funding acquisition. H. Steeb: Conceptualization, Writing - Review \& Editing, Supervision, Funding acquisition. M. Schneider: Conceptualization, Methodology, Investigation, Software, Validation, Writing - Original Draft, Visualization, Project administration.

\section*{Acknowledgements}
Funded by Deutsche Forschungsgemeinschaft (DFG, German Research Foundation) under Germany's Excellence Strategy (Project number 390740016 - EXC 2075) and the Collaborative Research Center 1313 (project number 327154368 - SFB1313). We acknowledge the support by the Stuttgart Center for Simulation Science (SC SimTech). Felix Weinhardt thanks the Alexander von Humboldt Foundation for supporting this work through a Feodor-Lynen Research Fellowship.

\bibliographystyle{elsarticle-num-names}
\bibliography{cas-refs}

\end{document}